\documentclass[showpacs,aps,prl,reprint,superscriptaddress,preprintnumbers,nolinenumbers]{revtex4-1}

\usepackage{amsmath}
\usepackage{amssymb}
\usepackage{graphicx}
\usepackage{dcolumn}
\usepackage[dvipsnames]{xcolor}
\usepackage{amsthm,amssymb}
\usepackage{mathrsfs}
\usepackage{color}
\usepackage{endnotes}
\usepackage{bm}
\usepackage{epsfig}
\usepackage{array}
\usepackage{ulem}
\usepackage{color,soul}

\newcommand{\Tast}{$T^{\rm \ast}$}
\newcommand{\etal}{{\it et al.}}
\newcommand{\ie}{{\it i.e.\ }}
\newcommand{\CsVSb}{{CsV$_{3}$Sb$_{5}$}}
\newcommand{\AVSb}{{AV$_{3}$Sb$_{5}$}} 
\newcommand{\KVSb}{{KV$_{3}$Sb$_{5}$}} 
\newcommand{\RbVSb}{{RbV$_{3}$Sb$_{5}$}} 
\newcommand{\Tc}{$T_{\rm c}$}
\newcommand{\Tcdw}{$T_{\rm CDW}$}

\begin{document}

\title{ Discovery of a new phase in thin flakes of KV$_{3}$Sb$_{5}$ under pressure}

\author{Zheyu~Wang}
\author{Lingfei~Wang}
\author{King~Yau~Yip}
\author{Ying~Kit~Tsui}
\affiliation{Department of Physics, The Chinese University of Hong Kong, Shatin, Hong Kong, China}
\author{Tsz~Fung~Poon}
\author{Wenyan~Wang}
\author{Chun~Wai~Tsang}
\affiliation{Department of Physics, The Chinese University of Hong Kong, Shatin, Hong Kong, China}
\author{Shanmin~Wang}
\affiliation{Department of Physics, Southern University of Science and Technology, Shenzhen, Guangdong, China}
\author{David~Graf}
\affiliation{National High Magnetic Field Laboratory, Florida State University, Tallahassee, FL, USA}
\author{Alexandre~Pourret}
\affiliation{Univ. Grenoble Alpes, CEA, Grenoble-INP, IRIG, Pheliqs, 38000 Grenoble, France}
\author{Gabriel~Seyfarth}
\affiliation{Univ. Grenoble Alpes, INSA Toulouse, Univ. Toulouse Paul Sabatier,
EMFL, CNRS, LNCMI, Grenoble 38042, France}
\author{Georg~Knebel}
\affiliation{Univ. Grenoble Alpes, CEA, Grenoble-INP, IRIG, Pheliqs, 38000 Grenoble, France}
\author{Kwing~To~Lai}
\affiliation{Department of Physics, The Chinese University of Hong Kong, Shatin, Hong Kong, China} 
\author{Wing~Chi~Yu}
\affiliation{Department of Physics, City University of Hong Kong, Kowloon, Hong Kong, China} 
\author{Wei~Zhang}
\email[]{wzhang@phy.cuhk.edu.hk}
\affiliation{Department of Physics, The Chinese University of Hong Kong, Shatin, Hong Kong, China} 
\author{Swee~K.~Goh}
\email[]{skgoh@cuhk.edu.hk}
\affiliation{Department of Physics, The Chinese University of Hong Kong, Shatin, Hong Kong, China}

\date{\today}

\begin{abstract}
We report results of magnetotransport measurements on KV$_3$Sb$_5$ thin flakes under pressure. Our zero-field electrical resistance reveals an additional anomaly emerging under pressure ($p$), marking a previously unidentified phase boundary $T^{\rm \ast}$($p$).  Together with the established $T_{\rm CDW}(p)$ and $T_c(p)$, denoting the charge-density-wave transition and a superconducting transition, respectively, the temperature-pressure phase diagram of KV$_3$Sb$_5$ features a rich interplay among multiple phases. The Hall coefficient evolves reasonably smoothly when crossing the $T^{\rm \ast}$ phase boundary compared with the variation when crossing $T_{\rm CDW}$, indicating the preservation of the pristine electronic structure. The mobility spectrum analysis provides further insights into distinguishing different phases. Finally, our high-pressure quantum oscillation studies up to 31~T combined with the density functional theory calculations further demonstrate that the new phase does not reconstruct the Fermi surface, confirming that the translational symmetry of the pristine metallic state is preserved. 

\end{abstract}

\maketitle

\noindent {\bf \large 1. Introduction}\\
\noindent The kagome lattice, a two-dimensional network of corner-sharing triangles, intrinsically accommodates Dirac fermions, van-Hove singularities (vHs) and flat bands~\cite{Neupert2021,Ortiz2019,Ortiz2020,Li2021,Jiang2023,Yin2022}. Due to the diverging density of states at the vHs and flat bands, a variety of correlated many-body ground states can be realized. In kagome metal \AVSb\ (A=K, Rb, Cs), both charge density wave order~\cite{Luo2022,Liang2021,HU2022,Zhao2021,Kato2023} (\Tcdw$\sim$80--100~K) and superconductivity~\cite{Ortiz2020,Ortiz2021a,Yin2021} (\Tc$\sim$0.9--2.7~K) have been revealed by various measurements. The charge density wave (CDW) phase has attracted significant attention for its unconventional characteristics, in particular, whether time-reversal symmetry breaking (TRSB) occurs within the CDW phase~\cite{Yang2020,Wang2023a,Yu2021b,Mielke2022,Guo2022,Saykin2023,Xu2022}. While the issue of TRSB remains under debate, the understanding of the CDW superlattice has gained substantial progress. Regarding in-plane distortion, both ``Star of David" and trihexagonal patterns can be adopted~\cite{Tan2021} and different stacking arrangements along $c$-direction are observed among \KVSb, \CsVSb\ and \RbVSb, indicating the 3D nature of the CDW order~\cite{Kang2022a, Ortiz2021, Liang2021}. Besides, a stripe order of CDW with a modulation of 4$a_0$ has been revealed by the detailed scanning tunneling microscopy (STM) measurement~\cite{Zheng2022} and a commensurate CDW to incommensurate CDW (ICCDW) transition under pressure has been proposed by the nuclear quadrupole resonance (NQR) measurement~\cite{Feng2023}. The interplay between CDW and superconductivity also presents intriguing dynamics. While the CDW order is suppressed monotonically as pressure increases, the double-dome feature of \Tc\ has been discovered across all three compounds~\cite{Zhu2022,Du2021,Chen2021a}, with \Tc\ reaching a maximum near the CDW suppression point, suggesting the potential role of charge fluctuations on superconductivity~\cite{Wang2024a}.

Despite the ongoing debate on the nature of the unconventional CDW and superconductivity, more recent studies have expanded the general interest to examine their interplay with other correlated phenomena, notably the possibility of electronic nematicity. While the elastoresistance~\cite{Sur2023}, nuclear magnetic resonance~\cite{Nie2022}, STM~\cite{Jiang2021,Li2022} and zero-field muon spin relaxation measurements~\cite{Yu2021c} have pointed towards the presence of the nematic order with $C_2$ symmetry inside the CDW phase, a recent high-resolution torque measurement in \CsVSb\ unveiled an odd-parity nematic phase~\cite{Asaba2024} above \Tcdw. Furthermore, in-plane transport measurements on \CsVSb\ and \RbVSb\ have uncovered a two-fold symmetry of the superconducting phase~\cite{Xiang2021,Wang2024}, indicating the intimate interplay between nematicity and superconductivity.

In this manuscript, we report the unexpected emergence of a new phase in \KVSb\ under pressure. The anomaly representing the transition temperature manifests in the transport data at low pressure and decreases monotonically as pressure increases. The relatively smooth evolution of the Hall coefficient when cooling across the phase boundary indicates the absence of a severe Fermi surface reconstruction, and the mobility spectrum analysis derived from the magnetotransport data offers a deeper insight to distinguish the new phase from other features. Finally, we present the quantum oscillation and density functional theory calculation results to confirm that the pristine Fermi surface of \KVSb\ is preserved in the new phase. Our result suggests that nematicity is the leading candidate of the new phase. The newly constructed temperature-pressure phase diagram uncovers an intricate interplay among the new phase, CDW and the superconductivity in the kagome system and reaffirms that kagome lattices provide a novel platform to stabilize novel phases that are often intertwined with superconductivity.       
\\

\begin{figure}[!t]\centering
      \resizebox{8.5cm}{!}{
              \includegraphics{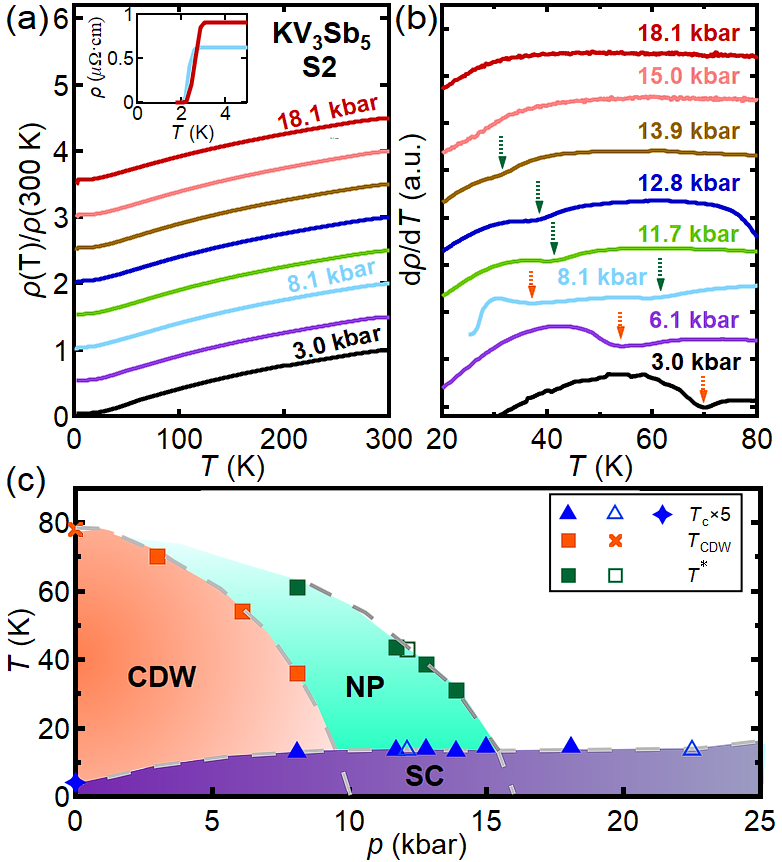}}              				
              \caption{\label{PhaDia}   
              Temperature dependence of  (a) the renormalized resistivity $\rho(T)/\rho({\rm 300~K})$ and (b) ${\rm d}\rho/{\rm d}T$ of S2 at different pressures. While the curves here are shifted vertically for clarity, data plotted in the absolute unit can be found in the Supporting Information~\cite{supp}. The inset in (a) displays the superconducting transition at 8.1~kbar and 18.1~kbar, while the orange solid arrows and the green dashed arrows in (b) respectively represent $T_{\rm CDW}$ and \Tast. For ${\rm d}\rho/{\rm d}T$ down to 10~K at 8.1~kbar, see Fig.~S6(c). (c) Temperature-pressure phase diagram of \KVSb. The solid symbols represent data from S2 and the open symbols are from S1. The star and the cross are ambient-pressure \Tc\ and \Tcdw\ taken from Ref.~\cite{Wang2023}, respectively. The $T_{\rm c}$ is multiplied by 5. }   
\end{figure}

\noindent {\bf \large 2. Experimental Results}\\
\noindent {\bf 2.1 Emergence of a pressure-induced phase in \KVSb\ thin flakes}\\
\noindent Figure~\ref{PhaDia}(a) shows the temperature dependence of the resistivity collected on S2 at various pressures, while the corresponding ${\rm d}\rho/{\rm d}T$ curves are presented in Fig.~\ref{PhaDia}(b). At 3.0~kbar and 6.1~kbar, dips in ${\rm d}\rho/{\rm d}T$ are recorded respectively at 70~K and 54~K (see orange arrows in Fig.~\ref{PhaDia}(b)). These features can be smoothly connected to the anomaly at 78~K observed in another thin flake reported in our earlier work at ambient pressure~\cite{Wang2023}, where we have established it to be a CDW phase transition. Thus, the anomaly at 3.0~kbar and 6.1~kbar are attributed to CDW phase transition temperatures $T_{\rm CDW}$. At 8.1~kbar, however, three clear features can be observed:  in addition to the superconducting transition at \Tc=2.6~K that can be unambiguously identified because of the zero resistance (see inset of Fig.~\ref{PhaDia}(a)), ${\rm d}\rho/{\rm d}T$ displays two minima instead of just one dip at higher temperatures, as indicated by the green and orange arrows in Fig.~\ref{PhaDia}(b), respectively. The dip at the lower temperature (36~K) extends the decreasing trend of $T_{\rm CDW}$ recorded at lower pressures. Therefore, it is natural to assign the anomaly at the lower temperature to $T_{\rm CDW}$. The anomaly at the higher temperature (61~K), on the other hand, have not been reported in the literature.

To learn more about the newly-observed anomaly at 8.1~kbar, we continue to increase the pressure. At 11.7~kbar, a clear minimum in ${\rm d}\rho/{\rm d}T$ is recorded at 43.5~K, between two minima at 8.1~kbar. Since $T_{\rm CDW}$ in all three AV$_3$Sb$_5$ systems have been reported to be a monotonically decreasing function of $p$ (Refs.~\cite{Chen2021a,Wang2021a,Yu2021}), it is unreasonable to assign the local minimum at 11.7~kbar to $T_{\rm CDW}$, as this would imply a sudden increase of $T_{\rm CDW}$ under pressure. Instead, this anomaly must be connected to the one observed at 8.1~kbar. It is now clear that a new temperature scale emerges in \KVSb\ under pressure, to which we attach a symbol $T^\ast$. Further increasing the pressure, a similar anomaly in ${\rm d}\rho/{\rm d}T$ can be followed to 13.9~kbar, enabling the construction of $T^\ast(p)$, which smoothly connects four datapoints in S2. At 15~kbar, no obvious dip can be discerned. 

Figure~\ref{PhaDia}(c) shows the temperature-pressure phase diagram constructed for thin flakes of \KVSb, including various anomalies detected.
 While $T_{\rm CDW}$ diminishes at lower pressures, \Tast\ persists to a higher pressure and extrapolates to 0~K at around 17~kbar. We stress that \Tast\ is observed in both S1 and S2: in S1, \Tast\ is 43~K at 12.1~kbar (open square in Fig.~\ref{PhaDia}(c)), in an excellent agreement with the pressure dependence of \Tast\ recorded in S2. Concomitant with the pressure evolution of $T_{\rm CDW}$ and $T^\ast$, $T_c$ with values greater than 2~K can also be observed for pressures greater than 8.1~kbar. Thus, Fig.~\ref{PhaDia}(c) is suggestive of an intriguing interplay among superconductivity, CDW and the new phase (NP) demarcated by \Tast.
\\

\begin{figure}[!t]\centering
      \resizebox{8.5cm}{!}{
 \includegraphics{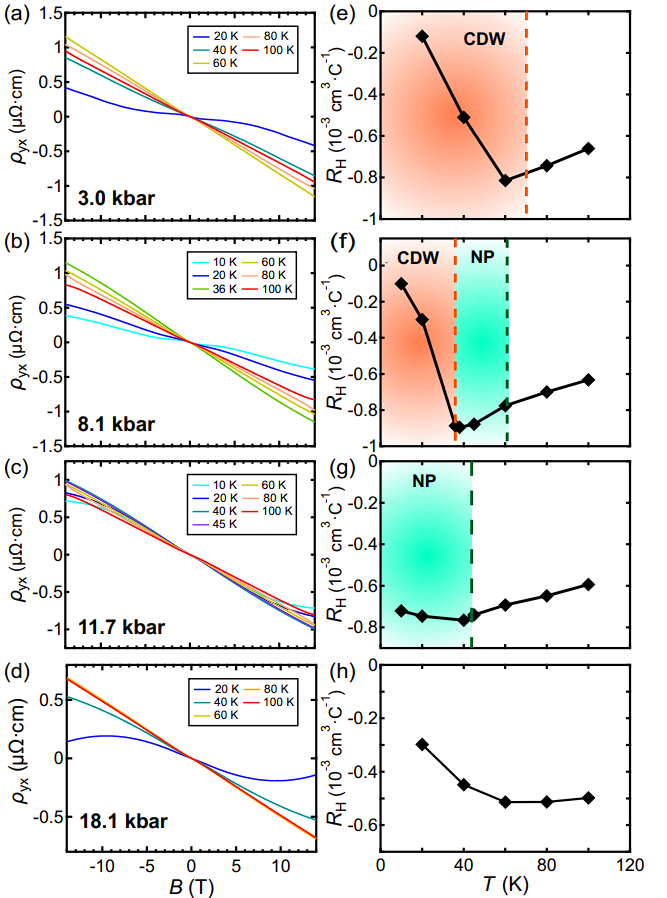}}       
 \caption{\label{Hall} Hall resistivity against the magnetic field at different temperatures for S2 at 3.0~kbar (a), 8.1~kbar (b), 11.7~kbar (c) and 18.1~kbar (d), respectively. Panels (e) to (h) depict the temperature dependence of the Hall coefficients at corresponding pressures. The orange and the green shaded areas represent the CDW phase and the \Tast-related phase respectively.}
          
\end{figure}

\noindent {\bf 2.2~Evolution of the Hall coefficient under pressure}\\
To investigate the new anomaly, we trace Hall signals of S2 at representative pressures and temperatures, depicted in Figs.~\ref{Hall}(a)-(d). At 3.0~kbar, where only $T_{\rm CDW}$ exists, a characteristic `S'-shape feature appears in the low-field region of $\rho_{\rm yx}$ at 20~K, but the feature diminishes above $T_{\rm CDW}$. This feature is consistent with the reported anomalous Hall effect (AHE) in \AVSb~\cite{Yang2020,Yu2021b,Zhang2023,Wang2023a}. At 80~K, $\rho_{\rm yx}$ is linear in field up to $\pm$14~T without the `S'-shape feature, indicating the system is in the pristine phase. Then we extract the Hall coefficient $R_{\rm H}$ from the slope of $\rho_{\rm yx}(B)$ at low field using a one-band approximation. For data distorted by the `S'-shape feature, $R_{\rm H}$ is calculated from the slope of $\rho_{\rm yx}(B)$ just beyond the `S'-shape region, following a procedure introduced in previous works on AV$_3$Sb$_5$ (see e.g. Ref.~[\onlinecite{Yang2020}]). As shown in Fig.~\ref{Hall}(e), $|R_{\rm H}|$ shows a maximum when crossing \Tcdw, and $|R_{\rm H}|$ significantly decreases with further cooling. Since the carrier density is inversely proportional to $|R_{\rm H}|$, the temperature dependence of $|R_{\rm H}|$ is consistent with the evolution of the carrier density reported at ambient pressure in Ref.~[\onlinecite{Yang2020}].

At 8.1~kbar, at which both \Tast\ and $T_{\rm CDW}$ are detected in the normal state, the `S'-shape feature can still be observed at 10~K and 20~K but is greatly weakened. Between \Tcdw~(36~K) and \Tast~(61~K), the Hall signals no longer host the `S'-shape feature. Instead, $\rho_{\rm yx}$ is linear up to 10~T, and only deviates slightly from linearity beyond 10~T. Above \Tast, $\rho_{\rm yx}$ shows linear dependence on field up to 14~T. As displayed in Fig.~\ref{Hall}(f), $|R_{\rm H}|$ extracted with the same procedure again shows a maximum near \Tcdw. Surprisingly, $R_{\rm H}$ changes reasonably smoothly when crossing \Tast\ compared with the variation when crossing \Tcdw, indicating that \Tast\ does not significantly affect the electronic structure.

Next, we increase the pressure to 11.7~kbar where only \Tast\ appears in the normal state. Instead of the `S'-shape feature, $\rho_{\rm yx}$ shows linear behavior up to 5~T and deviates from the linearity at higher fields at 10~K and 20~K. For $T>40$~K, $\rho_{\rm yx}$ shows perfect linear dependence on the field. We show the extracted $R_{\rm H}$ in Fig.~\ref{Hall}(g). Once again, $|R_{\rm H}|$ does not show a significant decrease across \Tast. Finally, we investigate S2 at 18.1~kbar where no anomaly appears in the normal state. From the low-field slope of $\rho_{\rm yx}(B)$ at various temperatures plotted in Fig.~\ref{Hall}(d), $R_{\rm H}$ is extracted and displayed in Fig.~\ref{Hall}(h). At 18.1~kbar, $R_{\rm H}$ shows a weak temperature dependence, and it is worth mentioning that the magnitude of changes in $R_{\rm H}$ in Fig.~\ref{Hall}(h) is comparable to the changes in $R_{\rm H}$ at 11.7~kbar at which \Tast\ exists.
Taking into account all these data, we conclude that the new feature associated with \Tast\ does not appear to modify the electronic structure.

\begin{figure*}[!t]\centering
       \resizebox{14cm}{!}{
              \includegraphics{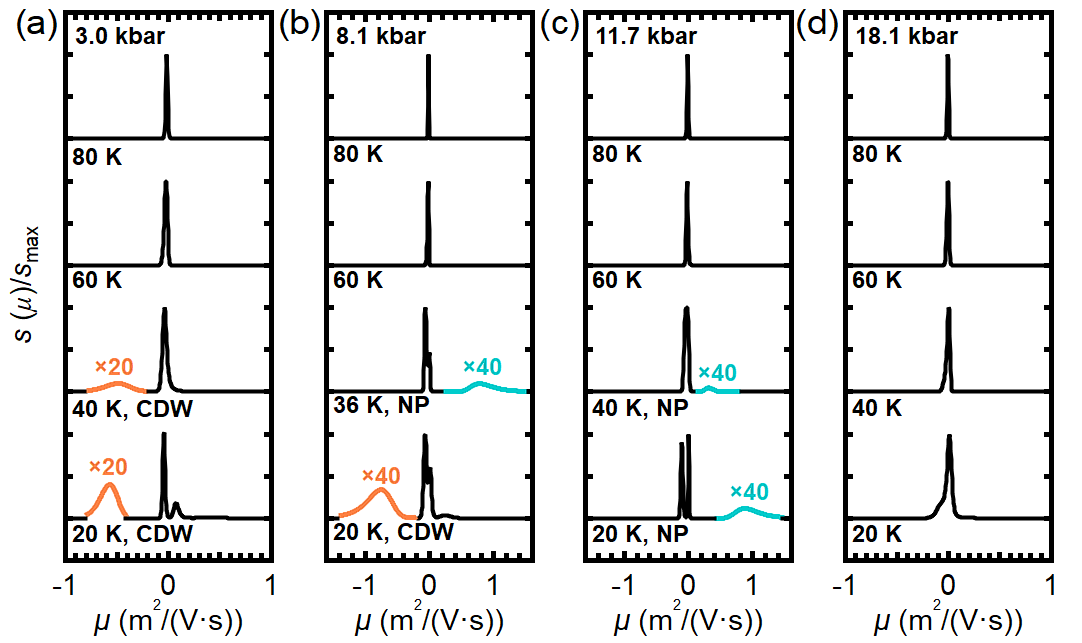}}                			
              \caption{\label{MSA} Mobility spectra of S2 at (a) 3.0~kbar, (b) 8.1~kbar, (c) 11.7~kbar and (d) 18.1~kbar. Each pressure includes mobility spectra at 80~K, 60~K, 40~K (except (b), where the spectrum at 36~K is shown) and 20~K with $s(\mu)$ renormalized as $s(\mu)/s_{\rm max}$, where $s_{\rm max}$ is the maximum value of $s(\mu)$ in each spectrum. Orange and green curves denote elements related to the CDW and \Tast-related phase, respectively, with their amplitudes magnified for clarity.
}
\end{figure*}

\vspace{5pt}
\noindent {\bf 2.3 Mobility spectrum as a diagnostic tool for phase transitions}\\
We further examine our transport data using the mobility spectrum analysis (MSA), which avoids any {\it a priori} assumptions about carrier types and numbers, and it provides detailed information on carrier mobilities~\cite{Kiatgamolchai2002,Beck1987}. The mobility spectra derived from the data on S2 are presented in Fig.~\ref{MSA}, covering a wide range of temperatures at four representative pressures. Additionally, a complete set of mobility spectra is plotted on the semilog scale in the Supporting Information~\cite{supp}. For the pristine phase at 18.1~kbar, the mobility spectrum changes from having only a single peak close to $\mu=0$ at high temperatures, to displaying multiple peaks at low temperatures. This evolution is consistent with previous reports that the high-temperature transport is dominated by one electron-like band while multiband transport becomes active at low temperatures~\cite{Yang2020,Yu2021b, Mi2022}. Other spectra in the pristine phase are broadly consistent with the data at 18.1~kbar. For example, the spectrum at 80~K and 11.7~kbar also exhibits a single peak near $\mu=0$.

New structures emerge in the mobility spectra when \Tast\ and \Tcdw\ start to play a role in transport. At 11.7~kbar, where \Tast\ is 43~K, a peak abruptly appears at around 0.3~${\rm m^2/Vs}$ when cooling to 40~K and further evolves to 0.9~${\rm m^2/Vs}$ at 20~K (Fig.~\ref{MSA}(c)). One possible interpretation is that a group of high-mobility holes emerge in the \Tast-related phase, and this set of holes has a significantly lower density, explaining the negligible effect introduced by \Tast\ on the electronic structure. Alternatively, the entrance into the \Tast-related phase introduces a subtle effect to $\rho_{\rm yx}(B)$, which manifests as a new peak in the mobility spectrum. To unambiguously settle this issue is difficult, because the new peak is very weak. If the positive peak were to manifest a group of holes, the Fermi surface should also undergo reconstruction upon entering the new phase. However, this contradicts the evidence from our quantum oscillation (as discussed later) and Hall resistivity results, which suggest no Fermi surface reconstruction at \Tast. Therefore, the positive peak is likely to have a different physical origin which requires further investigation. Nonetheless, this isolated peak structure provides a useful means to identify the \Tast-related phase. At 8.1~kbar, at which both \Tast\ and \Tcdw\ exist, new structures are observed in the mobility spectra. As shown in Fig.~\ref{MSA}(b), at 80~K and 60~K, a single-peak structure appears near $\mu=0$ as expected, indicating the system stays in the pristine phase. Cooling to 36~K, a group of high-mobility holes is found, signaling the stabilization of the \Tast-related phase. Further cooling to 20~K, which is lower than $T_{\rm CDW}$, a new peak suddenly appears on the negative mobility side, centered at around $-0.7~{\rm m^2/Vs}$. Phenomenologically, this new peak can be associated with the `S'-shape feature found in the Hall signals in the CDW phase, because the `S'-shape feature has a large, negative slope in the low field region, and the MSA routine interprets it as a group of high-mobility electrons. However, since the `S'-shape feature is recognized as the characteristic of the AHE in the \AVSb\ family~\cite{Yang2020,Yu2021b,Zhang2023,Wang2023a}, we adopt the mainstream interpretation that the new mobility peak is associated with the AHE~\cite{Jiang2021,Nagaosa2010}. Nevertheless, the high mobility peak on the negative side still serves as an indicator of the CDW phase. This indicator can also be found at 40~K and 20~K at 3.0~kbar (Fig.~\ref{MSA}(a)), where \Tcdw\ is 70~K. In conclusion, MSA provides diagnostic value for differentiating various phases in \KVSb, obtaining results agreeing with the $T$-$p$ phase diagram constructed using $\rho(T)$.

\begin{figure*}[!t]\centering
      \resizebox{13cm}{!}{
 \includegraphics{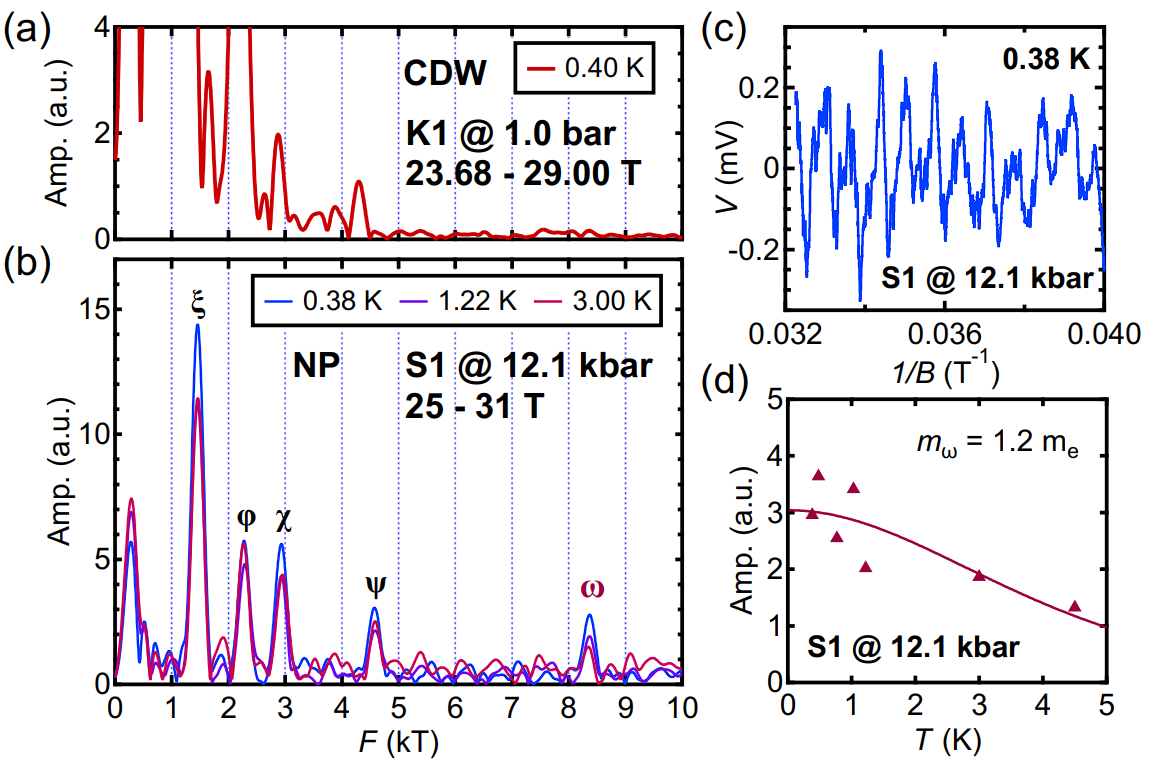}}       
 \caption{\label{QO} (a) FFT spectrum of the SdH oscillation data collected on K1 at 0.4~K at ambient pressure from 23.68~T to 29.00~T.(b) FFT spectra of the SdH oscillation data collected on S1 at 12.1~kbar from 25~T to 31~T at three typical temperatures. Peaks are marked with $\xi\sim$ 1453~T, $\varphi\sim$ 2277~T, $\chi$ $\sim$ 2930~T, $\psi\sim$ 4577~T and $\omega\sim$ 8372~T. The magnetic field is aligned along the $c$-axis when conducting measurements in (a) and (b). (c) Oscillations of resistive voltage collected on S1 at 0.38~K with the background subtracted. (d) Temperature dependence of the amplitudes of $\omega$ derived from the FFT spectra collected on S1. The solid curve corresponds to Lifshitz-Kosevich analysis.
}            
\end{figure*}

Finally, we discuss the apparent relationship between the nonlinearity in $\rho_{\rm yx}(B)$ at high field and the appearance of the high-mobility peak on the positive mobility side, and whether the nonlinearity can be used as an indicator for the \Tast-related phase. Our $\rho_{\rm yx}(B)$ appears to deviate from linearity at the high-field end below \Tast, see for example the dataset in the \Tast-related phase at 11.7 kbar. Since the MSA gives a high-mobility peak below \Tast, it might be tempting to link the two observations. However, one can reconstruct $\rho_{\rm yx}(B)$ using the mobility spectrum. By intentionally excluding the high-mobility peak, nonlinearity can still be observed in the reconstructed $\rho_{\rm yx}(B)$, consistent with experimental data. Full technical details are provided in the Supporting Information~\cite{supp}. Thus, the high-mobility peak on the positive side is not the cause of the nonlinearity in $\rho_{\rm yx}(B)$.

\vspace{5pt}
\noindent{\bf 2.4~Comparison between SdH oscillations and DFT calculation results}\\
To gain deeper insights into the new phase, we searched for clues from the perspective of electronic structures. We compare the electronic structure in the \Tast-related phase with the electronic structures in other phases in \KVSb. 

To begin with, we revealed the electronic structure of the CDW phase by performing high-field Shubnikov-de Haas (SdH) measurements on another thin flake sample (K1) up to 29~T and down to 0.4~K at ambient pressure, which undergoes a Fermi surface reconstruction at \Tcdw. The corresponding fast Fourier transform (FFT) analysis focusing on frequencies over 2000~T is shown in Fig.~\ref{QO}(a), while the spectrum exhibiting all frequencies is presented in the Supporting Information~\cite{supp}. The result is broadly consistent with the published data in Ref.~\cite{Wang2023} except the appearance of frequencies at around 2900~T and around 4300~T, which are considered as second harmonics as explained later. Notably, no frequency larger than $\sim$4300~T is recorded. Next, we explored the electronic structure of the \Tast-related phase by measuring the magnetoresistance up to 31~T in another pressure cell, in which we carefully increased the pressure of S1 to 12.1~kbar directly without going through any CDW phase transition. As described earlier, only the \Tast-related anomaly was detected at 12.1~kbar in the normal state of S1, and the value of \Tast\ agrees with $T^\ast(p)$ of S2 (Fig.~\ref{PhaDia}(c)). The FFT spectra of the SdH data from 25~T to 31~T are presented in Fig.~\ref{QO}(b), using the same horizontal scale as Fig.~\ref{QO}(a), which enables direct comparison between these two panels. In Fig.~\ref{QO}(c), we plot a representative oscillatory dataset (S1 at 12.1~kbar, 0.38~K) in the inverse field domain, the FFT of which is included in Fig.~\ref{QO}(b). In all high-field measurements above, the magnetic field was applied along the $c$-axis. Comparing S1 with K1, and S1 with the published data in Ref.~\cite{Wang2023}, we make two remarks about the spectrum at 12.1~kbar: (i) the FFT spectrum of the present study is significantly simpler, and (ii) a large frequency of 8372~T (marked with $\omega$), which is not revealed at ambient pressure, is observed. Finally, the corresponding quasi-particle effective mass of $\omega$ is estimated by Lifshitz-Kosevich analysis, as shown in Fig.~\ref{QO}(d).

\begin{figure}[!t]\centering
      \resizebox{8.5cm}{!}{
 \includegraphics{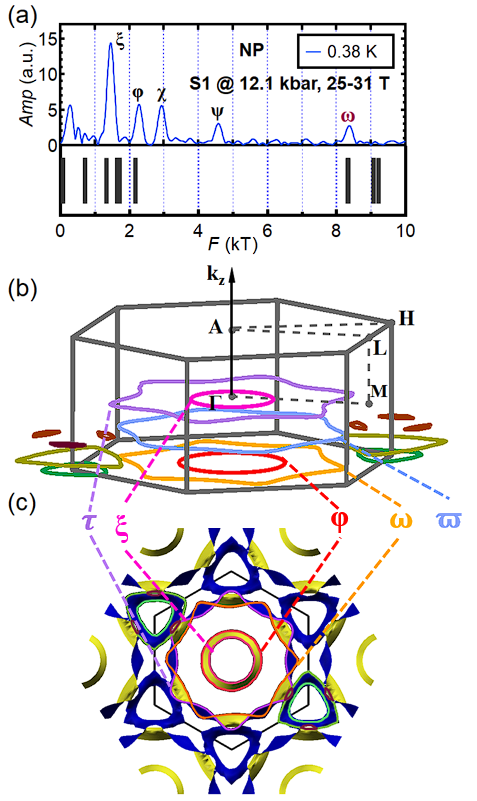}}       
 \caption{\label{DFT} (a) Comparison between the SdH frequencies at 12.1~kbar and 0.38~K of S1 (upper panel) and the frequencies of Band 124 proposed by DFT calculations in the undistorted pristine phase (lower panel). The magnetic field is aligned along the $c$-axis. (b) Calculated extremal orbits parallel to the $k_{\rm x}$-$k_{\rm _y}$ plane for Band 124. Only orbits with $k_{\rm z}\leq$ 0 are plotted for clarity. (c) Top views of the Fermi surfaces of Band 124 with the extremal orbits superimposed. The same color scheme is used for the extremal orbits in (b) and (c).
}            
\end{figure}

The SdH frequency at 8372~T is reminiscent of the large extremal orbit revealed recently by quantum oscillations in CsV$_3$Sb$_5$. In \CsVSb, a large SdH frequency ($\sim$8200~T) has recently been detected in the high-pressure `pristine' metallic state, \ie when the CDW transition is tuned away~\cite{Zhang2024}. The CDW order introduces an in-plane superlattice modulation resulting in a unit cell that is ($2\times2$) times larger. Consequently, the Brillouin zone area in the $k_x$-$k_y$ plane is reduced to approximately 4100~T for \AVSb. In both cases, the reconstructed Brillouin zone areas are too small to host an in-plane extremal orbit corresponding to a frequency as large as 8~kT. Thus, the detection of such large frequencies can be regarded as an indicator for the absence of the CDW order. Hence, it is tempting to assign the observed 8372~T frequency in \KVSb\ at 12.1~kbar to the large Fermi surface of the pristine metallic state, akin to CsV$_3$Sb$_5$, even though at 12.1~kbar the \Tast-related phase is present.

To investigate the fermiology of the pristine metallic state, we perform density functional theory calculations. The experimental data at 12.1~kbar and calculated frequencies are presented in Fig.~\ref{DFT}(a). In Fig.~\ref{DFT}(b), the extremal orbits at various $k_z$ are shown. We also show the top views of the Fermi surfaces within the first Brillouin zone in Fig.~\ref{DFT}(c). Interestingly, the calculated results display a good agreement with the experimentally detected SdH frequencies at 12.1~kbar, despite the fact that the calculations do not assume any symmetry-lowering order. 
For peaks labelled by ${\rm \xi}$, ${\rm \varphi}$, and ${\rm \omega}$, the agreement between the experimental results and calculations is particularly good (Fig.~\ref{DFT}(a)). The corresponding calculated extremal orbits can be seen in Figs.~\ref{DFT}(b) and (c), with $\omega$ occupying nearly 52~\% of the Brillouin zone area. The orbits labelled by $\tau$ and $\varpi$, which exceed 9000~T according to calculations, are not observed in our experiments. We hope that by enhancing the signal-to-noise ratio further, these frequencies can be detected in the future. No calculated values can be assigned to $\chi$ and $\psi$. Given that $\chi$ and $\psi$ approximate twice the values of $\xi$ and $\varphi$, they could be the second harmonics. Through the thermal damping factor of the Lifshitz-Kosevich theory, the quasiparticle effective masses ($m^*$) can be calculated (see Fig.~S4). Unfortunately, the relatively weak amplitudes prevent a conclusive verification of the harmonic relationship through the values of $m^*$. Overall, the good agreement between the DFT calculation in the pristine phase and the experimental results in the \Tast-related phase emphasizes that the Fermi surface is not reconstructed in the new phase.

\vspace{5pt}
\noindent {\bf \large 3. Discussion}\\
\noindent  We now discuss the nature of the \Tast-related phase based on existing experimental data. An evolution of the CDW phase from a commensurate long-range order to an incommensurate order under pressure has been proposed by nuclear quadrupole resonance (NQR)~\cite{Feng2023} in \CsVSb. It is conceivable that \KVSb\ also enters an ICCDW order at \Tast, but \Tast\ has a different pressure dependence from the commensurate CDW order denoted by \Tcdw. Such a temperature-pressure phase diagram has been constructed for 1T-TaS$_2$ (Refs.~\cite{Sipos2008} and \cite{Perfetti2006}). However, it remains difficult to explain why \Tast\ is not observed at ambient pressure, although it could be very close to \Tcdw\ at low pressures such that it is hard to distinguish them. Importantly, the Fermi surface is expected to be reconstructed across the incommensurate CDW transition too -- this has not been observed experimentally, based on our Hall data and quantum oscillations. 

Another possibility arises from the spin degrees of freedom. Very recently, the coexistence of SDW phase, CDW phase and superconductivity has been observed in the Cr-based kagome metal ${\rm CsCr_3Sb_5}$ under pressure~\cite{Liu2024}. The reported $T$-$p$ phase diagram is remarkably similar to the present case, thus it is reasonable to postulate an SDW transition at \Tast\ in \KVSb.  However, at ambient pressure the metallic states preceding the superconductivity are different in ${\rm CsCr_3Sb_5}$ and \KVSb. In ${\rm CsCr_3Sb_5}$, an antiferromagnetic (AFM) state has been confirmed by nuclear magnetic resonance~\cite{Liu2024, Xu2023}. In \KVSb, however, evidence is accumulating that long-range magnetic order does not exist~\cite{Ortiz2019,Mielke2022,song2021a}. In particular, muon spin spectroscopy ($\mu$-SR)  reported the absence of local moments in \KVSb~\cite{Kenney2021}.  Because of the multiband nature of \KVSb, an SDW order should reconstruct the complicated Fermi surface, which has not been observed in the Hall data and quantum oscillations. Thus, the SDW is also unlikely to be the \Tast-related phase.

The need to preserve the pristine Fermi surface in the \Tast-related phase imposes a strong constraint on the possible nature of the electronic state. A leading candidate is an electronic nematicity, which has been proposed in \AVSb\ compounds~\cite{Xiang2021,Nie2022,Sur2023,Asaba2024, Jiang2023}. Nematicity breaks rotational symmetry but preserves translational symmetry. The fact that \Tast\ has not been observed in bulk \KVSb\ (Ref.~\cite{Du2021}) but appears in our thin flakes lends support to this scenario, because in the bulk sample multiple nematic domains can weaken signals from experimental probes that lack spatial resolution, due to the averaging effect. In our experiment, not only the thickness is significantly reduced but the separation between the voltage leads is also small ({10~$\mu$m}), leading to the possibility of probing only a dominant nematic domain in which the nematic order parameter is uniform. The relation between the sample size and the nematicity is explored earlier in \CsVSb~\cite{Asaba2024}, in which the two-fold torque signals were detected only in small samples below $\sim$100~K.

In \KVSb, one can envisage a distortion of the Fermi surface from the perfect six-fold symmetry, thereby breaking the rotational symmetry. While the shape of the Fermi surface is deformed in this manner, the cross-sectional area can be preserved. Thus, the quantum oscillation frequency does not vary and certainly does not exhibit a discontinuity. Such a smooth evolution of the quantum oscillation frequency across a nematic phase transition has been observed. For instance, when FeSe is driven across the nematic phase transition by sulfur substitution, the largest extremal orbit $\delta$ evolves smoothly and changes from an ellipse (rotational symmetry broken) in the nematic phase to a circle (rotational symmetry restored) outside the nematic region~\cite{Coldea2019}. Such a scenario, proposed by DFT and confirmed by quantum oscillations for the case of FeSe~(Ref.~\cite{Coldea2019}), is consistent with our observation for \KVSb. 

In conclusion, we discover a new phase in thin flakes of \KVSb\ and trace its evolution through magnetotransport. The new phase emerges at low pressure and persists beyond the pressure at which the CDW phase is completely suppressed. The Hall measurements imply the absence of a severe Fermi surface reconstruction when cooling across the phase boundary. 
The comparison between quantum oscillation results and DFT calculations further confirms the preservation of the pristine metallic Fermi surface in the new phase. Based on our experimental evidence, we tentatively attribute the new phase to nematicity. Given that both the newly discovered phase and the CDW phase are in proximity to the superconducting phase, it is crucial to confirm the broken symmetry associated with the new phase to understand the dome-shaped dependence of the superconducting transition temperature. It is also pressing to investigate the role of pressure in tuning the new phase, and explore the existence of this phase in sister compounds \RbVSb\ and \CsVSb. The discovery of the new phase in \KVSb\ provides a prime confirmation that the kagome lattice is a prolific platform for realizing multiple exotic phases.
 \\

\noindent {\bf \large 4. Experimental Section}\\
\noindent {\bf 4.1 Sample Preparation and Measurements}\\
\noindent Single crystals of \KVSb\ used in this work were synthesized using the self-flux method as described in  Ref.~\cite{Wang2023}. Three \KVSb\ thin flakes with similar thicknesses of around 200~nm, S1, S2 and K1 were cleaved from bulk single crystals from the same batch. While high-pressure magnetotransport measurements of S1 and S2 were conducted with a device-integrated diamond anvil cell (DIDAC) method~\cite{Xie2021, Zhang2023, Ku2022}, K1 was measured at ambient pressure. Both S1 and S2 were measured in a Physical Property Measurement System (PPMS) by Quantum Design, which provides temperature down to 2~K and magnetic field up to 14~T, while additional studies were conducted on S1 down to 0.38~K and up to 31~T at National High Magnetic Field Laboratory (NHMFL) in Tallahassee. Finally, K1 was studied down to 0.4~K and up to 29~T at Laboratoire National des Champs Magnétiques Intenses (LNCMI) in Grenoble. \\
\\
\\
\noindent {\bf 4.2 Mobility Spectrum Analysis}\\
The mobility spectrum analysis (MSA)~\cite{Beck1987,Kiatgamolchai2002} has been conducted by combining both transverse magnetoresistivity and conventional Hall resistivity. The magnetotransport data were collected in S2 with the magnetic field applied along the $c$ direction. From experiments, we obtain the resistivity tensor components $\rho_{xx}$ and $\rho_{yx}$. They are then converted into conductivity tensor components $\sigma_{xx}$ and $\sigma_{xy}$. MSA takes $\sigma_{xx}$ and $\sigma_{xy}$, and outputs the mobility spectrum. The spectrum attributes the transport to the presence of carriers of different mobilities. The MSA method has been successfully applied on FeSe and PtBi$_2$. For FeSe, MSA enabled the detection of the high-mobility electrons~\cite{Huynh2014,Liam2022}. For PtBi$_2$, MSA confirmed the presence of five bands in agreement with first principle calculation and ARPES measurement~\cite{Zhao2021B}. There are multiple approaches to the MSA~\cite{Beck2021}. In this paper, we have adopted the maximum entropy approach introduced by Kiatgamolchai~\etal~\cite{Kiatgamolchai2002}.\\

\noindent {\bf 4.3 Density Functional Theory Calculations}\\
The WIEN2k package~\cite{Schwarz2003}, which employs density functional theory (DFT) with full-electron full-potential linearized augmented plane waves, was used in the Fermi surface calculation. The lattice constants of \KVSb\ were adopted from Ref.~\cite{Tan2021} and no structural optimization was performed. Further details of the DFT calculation can be found in the Supporting Information S1. 
\\
\\
\noindent {\bf  Supporting Information}\\
Supporting Information is available from the Wiley Online Library or from the author.\\

\begin{acknowledgments}
\noindent {\bf Acknowledgements}\\
\noindent We acknowledge Ziqiang Wang for fruitful discussions. 
The work was supported by Research Grants Council of Hong Kong (A-CUHK 402/19, CUHK 14301020, CUHK 14300722, CUHK 14302724), CUHK Direct Grant (4053577, 4053525), City University of Hong Kong (9610438), French National Agency for Research (ANR) within the project FETTOM (ANR-19-CE30-0037), the National Natural Science Foundation of China (Grant No. 12174175, 12104384), the Guangdong Provincial Quantum Science Strategic Initiative (GDZX2301009) and the Guangdong Basic and Applied Basic Research Foundation (Grant No. 2022B1515120014). A portion of this work was performed at the National High Magnetic Field Laboratory, which is supported by National Science Foundation Cooperative Agreement No. DMR-2128556 and the State of Florida. We also acknowledge the support of the LNCMI-CNRS, a member of the European Magnetic Field Laboratory (EMFL). \\ 
\end{acknowledgments}

\noindent {\bf  Conflict of Interest}\\
The authors declare no conflict of interest.\\

\noindent {\bf  Data Availability Statement}\\
The data that support the findings of this study are available from the corresponding author upon reasonable request.\\

\noindent {\bf  Keywords}\\
kagome superconductors, pressure-induced phase, mobility spectrum analysis, quantum oscillations


\providecommand{\noopsort}[1]{}\providecommand{\singleletter}[1]{#1}%

\end{document}